\documentclass[12pt]{article}
\usepackage{latexsym}
\newcommand{\be}{\begin{equation}}
\newcommand{\ee}{\end{equation}}
\def\n{\noindent}
\catcode `\@=11
\catcode `\@=12
\begin{document}
\begin{center}

{\bf {PLANE SYMMETRIC INHOMOGENEOUS BULK VISCOUS DOMAIN WALL IN LYRA GEOMETRY}} \\
\vspace{5mm}
\normalsize{ANIRUDH PRADHAN $^{\ast}$\footnote{Corresponding Author}, VANDANA RAI $^{\dag}$ 
and SAEED OTAROD {$^{\ddag}$}} \\
\vspace{10mm}
\normalsize{$^{\ast}$\it{Department of Mathematics, Hindu Post-graduate College,
Zamania-232 331, Ghazipur, India \\
E-mail : pradhan@iucaa.ernet.in, acpradhan@yahoo.com}}\\
\vspace{5mm}
\normalsize{$^{\dag}$\it{ Department of Mathematics, Post-graduate College,
Ghazipur-233 001, India}} \\
\vspace{5mm}
\normalsize{$^{\ddag}$\it{Department of Physics, Yasouj University,
Yasouj, Iran \\
E-mail : sotarod@mail.yu.ac.ir, sotarod@yahoo.com}}\\
\end{center}
\vspace{10mm}
\begin{abstract} 
Some bulk viscous general solutions are found for domain walls in Lyra geometry
in the plane symmetric inhomogeneous spacetime. Expressions for the energy density 
and pressure of domain walls are derived in both cases of uniform and time varying 
displacement field $\beta$. The viscosity coefficient of bulk viscous fluid is assumed 
to be a power function of mass density. Some physical consequences of the models are 
also given. Finally, the geodesic equations and acceleration of the test particle are 
discussed.
\end{abstract}
\smallskip
\n PACS: {98.80.-k, 75.60.Ch} \\ 
\smallskip
\n Keywords: {cosmology,  plane symmetric domain walls, bulk viscous model, 
Lyra geometry}
\section{Introduction}
Topological structures could be produced at phase transitions in the universe 
as it cooled {\cite{ref1}$-$\cite{ref5}}. Phase transitions can also give birth 
to solitonlike structures such as monopoles, strings and domain wall\cite{ref6}.
Within the context of general relativity, domain walls are immediately recognizable 
as especially unusual and interesting sources of gravity. Domain walls form when 
discrete symmetry is spontaneously broken\cite{ref7}. In simplest models, symmetry 
breaking is accomplished by a real scalar field $\phi$
whose vacuum manifold is disconnected. For example, suppose that the scalar 
potential for $\phi$ is $U(\phi) = \lambda (\phi^{2} - \mu^{2})^2$. The vacuum 
manifold for $\phi$ then consists of the two points [$\phi = \mu $ and 
$\phi = - \mu$]. After symmetry breaking, different regions of the universe can 
settle into different parts of the vacuum with domain walls forming the boundaries 
between these regions. As was pointed out by Zel'dovich {\it et al.}\cite{ref6}, the 
stress-energy of domain walls is composed of surface energy density and strong 
tension in two spatial directions, with the magnitude of the tension equal to that 
of the surface energy density. This is interesting because there are several 
indications that tension acts as a repulsive source of gravity in general relativity 
whereas pressure is attractive. We note, however that this analysis 
neglects the effects of gravity\cite{ref8}. Locally, the stress energy for a wall 
of arbitrary shape is similar to that of a plane-symmetric wall having both surface 
energy density and surface tension. Closed-surface domain walls collapse due to 
their surface tension. However, the details of the collapse for a wall with 
arbitrary shape and finite thickness are largely unknown.

The spacetime of cosmological domain walls has now been a subject of interest for more 
than a decade since the work of Vilenkin\cite{ref9} and Ipser and Sikivie\cite{ref10}
who use Israel's thin wall formalism\cite{ref11} to compute the gravitational field of
an infinitesimally thin planar domain wall. After the original work\cite{ref9,ref10} for 
thin walls, attempts focused on trying to find a 
perturbative expansion in the wall thickness\cite{ref8,ref12}. With the proposition by 
Hill, Schramn and Fry\cite{ref13} of a late phase transition with thick domain walls, 
there were some effort in finding exact thick solution\cite{ref14,ref15}. Recently, 
Bonjour {\it et al.}\cite{ref16} considered gravitating thick domain wall solutions with 
planar and reflection symmetry in the Goldstone model. Bonjour {\it et al.}\cite{ref17} 
also investigated the spacetime of a thick gravitational domain wall for a general 
potential $V(\phi)$. Jensen and Soleng \cite{ref18} have studied anisotropic domain 
walls where the solution has naked singularities and the generic solution is unstable 
to Hawking decay.

The investigation of relativistic cosmological models usually has the energy momentum
tensor of matter generated by a perfect fluid. To consider more realistic models one must
take into account the viscosity mechanisms, which have already attracted the attention
of many researchers. Most of the studies in cosmology involve a perfect fluid. Large entropy 
per baryon and the remarkable degree of isotropy of the cosmic microwave background 
radiation, suggests that we should analyze dissipative effects in cosmology. Further,
there are several processes which are expected to give rise to viscous effect. 
These are the decoupling of neutrinos during the radiation era and the recombination 
era\cite{ref19}, decay of massive super string modes into massless modes \cite{ref20},
gravitational string production\cite{ref21,ref22} and particle creation effect in 
grand unification era. It is known that the introduction of bulk 
viscosity can avoid the big bang singularity. Thus, we should consider the presence 
of a material distribution other than a perfect fluid to have realistic 
cosmological models (see Gr\o n\cite{ref23} for a review on cosmological 
models with bulk viscosity). A uniform cosmological model filled with fluid which possesses
pressure and second (bulk) viscosity was developed by Murphy\cite{ref24}. The solutions that
he found exhibit an interesting feature that the big bang type singularity appears in the 
infinite past.  

Einstein (1917) geometrized gravitation. Weyl, in 1918, was inspired by it and he was the the 
first to unify gravitation and electromagnetism in a single spacetime geometry. He showed 
how can one introduce a vector field in the Riemannian spacetime with an intrinsic geometrical 
significance. But this theory was not accepted as it was based on non-integrability of length 
transfer. Lyra\cite{ref25} introduced a gauge function, i.e., a displacement vector in 
Riemannian spacetime which removes the non-integrability condition of a vector under 
parallel transport. In this way Riemannian geometry was given a new modification by him 
and the modified geometry was named as Lyra's geometry. 

Sen\cite{ref26} and Sen and Dunn\cite{ref27} have proposed a new scalar-tensor theory of gravitation 
and constructed the field equations analogous to the Einstein's field equations, based on Lyra's 
geometry which in normal gauge may be written in the form 
\begin{equation}
\label{eq1}
R_{ij} - \frac{1}{2} g_{ij} R + \frac{3}{2} \phi_i \phi_j - \frac{3}{4} g_{ij} 
\phi_k \phi^k = - 8 \pi G T_{ij}, 
\end{equation}
\noindent 
where $\phi_{i}$ is the displacement vector and other symbols have their usual meanings. 

Halford\cite{ref28} has pointed out that the constant vector displacement
field $\phi_i$ in Lyra's geometry plays the role of cosmological
constant $\Lambda$ in the normal general relativistic treatment. It
is shown by Halford\cite{ref29} that the scalar-tensor treatment based on
Lyra's geometry predicts the same effects, within observational limits
as the Einstein's theory. Several investigators{\cite{ref30}$-$\cite{ref43}} have studied 
cosmological models based on Lyra geometry in different context. Soleng\cite{ref31} has 
pointed out that the cosmologies based on Lyra's manifold with constant gauge vector $\phi$ 
will either include a creation  field and be equal to Hoyle's creation field cosmology 
{\cite{ref44}$-$\cite{ref46}} or contain a special vacuum field which together with 
the gauge vector term may be considered as a cosmological term. In the latter case 
the solutions are equal to the general relativistic cosmologies with a cosmological term. 

The universe is spherically symmetric and the matter distribution in
it is on the whole isotropic and homogeneous. But during the early stages of 
evolution, it is unlikely that it could have had such a smoothed out picture. 
Hence, we consider plane symmetry which provides an opportunity for the study 
of inhomogeneity. Recently Pradhan {\it et al.}\cite{ref47} have studied plane symmetric 
domain wall in presence of a perfect fluid. 

Motivated by the situations discussed above, in this paper we shall focus upon the problem 
of establishing a formalism for studying the general solutions for domain wall in Lyra 
geometry in the plane symmetric inhomogeneous spacetime metric in presence of bulk viscous 
fluid. Expressions for the energy density and pressure of domain walls are obtained in 
both cases of uniform and time varying displacement field $\beta$. This paper is organized 
as follows: The metric and the basic equations are presented in Section 2. In Section 3 
we deal with the solution of field equations. The Subsection 3.1 contains the solution of 
uniform displacement field ($\beta = \beta_{0}$, constant). This section also contains 
two different models and also the physical consequences of these models. The 
Subsection 3.2 deal with the solution with time varying displacement field 
($\beta = \beta_{0}t^{\alpha}$). This subsection also contains two different models and 
their physical consequences are discussed. The geodesic equations and accelerations of 
the test particle are discussed in Section 4. Finally in Section 5 concluding remarks 
are given.  
\section{The metric and basic equations}
Thick domain walls are characterized by the energy momentum tensor of a viscous field
which has the form 
\begin{equation}
\label{eq2}
T_{i k} = \rho ( g_{ik} + w_{i}w_{k}) + \bar{p} w_{i} w_{k}, \; \;  w_{i} w^{i} = - 1,  
\end{equation}
\noindent 
where 
\begin{equation}
\label{eq3}
\bar{p} = p - \xi w^{i}_{;i}.
\end{equation}
Here $\rho$, $p$, $\bar{p}$, and $\xi$ are the energy density, the pressure in the 
direction normal to the plane of the wall, effective pressure, bulk viscous coefficient
respectively and $w_{i}$ is a unit space like vector in the same direction. 

The displacement vector $\phi_{i}$ in Equation (\ref{eq1}) is given by
\begin{equation}
\label{eq4}
\phi_{i} = (0, 0, 0, \beta),
\end{equation}
where $\beta$ may be considered constant as well as function of time coordinate like 
cosmological constant in Einstein's theory of gravitation. 

The energy momentum tensor $T_{ij}$ in comoving coordinates for thick domain walls take 
the form
\begin{equation}
\label{eq5}
T^{0}_{0} = T^{2}_{2} = T^{3}_{3}= \rho,~ ~  T^{1}_{1} = - \bar{p},~ ~  T^{0}_{1} = 0.
\end{equation}
\noindent 
We consider the most general plane symmetric spacetime metric suggested 
by Taub \cite{ref48} 
\begin{equation}
\label{eq6}
ds^2 = e^A (dt^2 - dz^2) - e^B (dx^2 + dy^2),
\end{equation}
\noindent 
where $A$ and $B$ are functions of $t$ and $z$. 

Using equation (\ref{eq5}) the field equations (\ref{eq4}) for the metric (\ref{eq6}) reduce to
\begin{equation}
\label{eq7}
\frac{e^{-A}}{4}(-4B'' - 3B'^2 + 2A' B') + \frac{e^{-A}}{4} (\dot B^2 + 2 \dot B\dot A)
- \frac{3}{4} e^{-A} \beta^2 = 8 \pi \rho,
\end{equation}
\begin{equation}
\label{eq8}
\frac{e^{-A}}{4}(- B'^2 - 2 B' A') + \frac{e^{-A}}{4}(- 4 \ddot B + 3 \dot B^2 - 2 \dot A 
\dot B) + \frac{3}{4} e^{-A} \beta^2 = - 8 \pi \bar{p},
\end{equation}
\begin{equation}
\label{eq9}
\frac{e^{-A}}{4}[-2(A'' + B'') - B'^2] + \frac{e^{-A}}{4}[2(\ddot A + \ddot B) + \dot B^2] 
+ \frac{3}{4} e^{-A} \beta^2 = 8 \pi \rho,
\end{equation}
\begin{equation}
\label{eq10}
- \dot {B}' + \dot B(A' - B') + \dot A B' = 0.
\end{equation}
In order to solve the above set of field equations we assume the separable form of
the metric coefficients as follows
\begin{equation}
\label{eq11}
A = A_1(z) + A_2(t),~ ~ ~ B = B_1(z) + B_2(t).
\end{equation}
From Eqs. (\ref{eq10}) and (\ref{eq11}), we obtain
\begin{equation}
\label{eq12}
\frac{A'_{1}}{B'_{1}} = \frac{(\dot B_2 -\dot A_2)}{B_2} = m,
\end{equation}
\noindent
where $m$ is considered as separation constant. 

Eq. (\ref{eq12}) yields the solution
\begin{equation}
\label{eq13}
A_1 = m B_1,
\end{equation}
\begin{equation}
\label{eq14}
A_2 = (1 - m) B_2.
\end{equation}
Again, subtracting Eq. (\ref{eq9}) from Eq. (\ref{eq7}) and using Eq. (\ref{eq11}), 
we obtain
\begin{equation}
\label{eq15}
A''_{1} - B''_{1} - B'^2_{1} + A'_{1} B'_{1} = \ddot A_{2} + \ddot B_{2} - \dot A_{2} 
\dot B_{2} + 3 \beta^2 = k,
\end{equation}
\noindent
where $k$ is another separation constant.

With the help of Eqs (\ref{eq13}) and (\ref{eq14}), Eq. (\ref{eq15}) may be written as
\begin{equation}
\label{eq16}
(m - 1)[B''_{1} + B'^2_{1}] = k,
\end{equation}
\begin{equation}
\label{eq17}
(2 - m) \ddot B_{2} + (m - 1) \dot B^2_{2} = k - 3 \beta^2.
\end{equation}
\section{Solutions of the field equations}
In this section we shall obtain exact solutions for thick domain walls in different
cases.

Using the substitution $u = e^{B_1}$ and $a = \frac{k}{1- m}$, Eq. (\ref{eq16}) 
takes the form
\begin{equation}
\label{eq18}
u'' + au = 0,
\end{equation}
which has the solution
\begin{equation}
\label{eq19}
e^{B_{1}} = u = c_{1}\sin(z\sqrt{a}) + c_{2}\cos(z\sqrt{a})  \; \; \; \mbox { when $a>0$},  
\end{equation}
where $c_1$ and $c_2$ are integrating constants. Eq. (\ref{eq19}) represent the 
general solution of the differential Eq. (\ref{eq18}) when $a > 0$. It may be noted that 
Rahaman {\it et al.}\cite{ref37} have obtained a particular solution for the case 
$a < 0$ in presence of perfect fluid. Recently Pradhan {\it et al.}\cite{ref47} have 
investigated a general solution in presence of perfect fluid.  

Eq. (\ref{eq17}) may be written as
\begin{equation}
\label{eq20}
\ddot{ B_{2}} - \frac{( 1- m)}{(2 - m)}\dot{ B^{2}_{2}} + \frac{3}{(2 - m)} \beta^2 = 
\frac{k}{2 - m}.
\end{equation}
Now we shall consider uniform and time varying displacement field $\beta$ separately.
\subsection{Case I: Uniform displacement field ($\beta = \beta_0$, constant)} 
By use of the transformation $v = e^{-\frac{(1 - m)}{(2 - m)}B_2}$, Eq. (\ref{eq19}) 
reduces to
\begin{equation}
\label{eq21}
\ddot v + b v = 0,
\end{equation}
where $$b = \frac{(1 - m) (k - 3\beta^2_0)}{(2 - m)^2}.$$
Again, it can be easily seen that Eq. (\ref{eq21}) possesses the solution
\begin{equation}
\label{eq22}
e^{-\frac{(1 - m)}{(2 -m)} B_2} = v = \bar{c_{1}}\sin(t\sqrt{b}) + \bar{c_{2}}\cos(t\sqrt{b}) 
\; \; \;  \mbox { when $b>0$}, 
\end{equation}
where $\bar{c_{1}}$ and $\bar{c_{2}}$ are integrating constants. Hence the metric coefficients 
have the explicit forms when $a>0$, $b>0$
\begin{equation}
\label{eq23} 
e^A = [c_{1}\sin(z\sqrt{a}) + c_{2}\cos(z\sqrt{a})]^{m}\times [\bar{c_{1}}\sin(t\sqrt{b}) 
+ \bar{c_{2}}\cos(t\sqrt{b})]^{(m - 2)},  
\end{equation}
\begin{equation}
\label{eq24}
e^B = [c_{1}\sin(z\sqrt{a}) + c_{2}\cos(z\sqrt{a})] \times [\bar{c_{1}}\sin(t\sqrt{b}) 
+ \bar{c_{2}}\cos(t\sqrt{b})]^{-\frac{(m - 2)}{(1 - m)}}.
\end{equation}
With the help of Eqs. (\ref{eq23}) and (\ref{eq24}), the energy density and pressure can 
be obtained from Eqs. (\ref{eq7}) and (\ref{eq8}) as given by
\begin{equation}
\label{eq25} 
32 \pi \rho = e^{-A}\left[4a + a\left(\frac{Z_1}{Z_2}\right)^2(1 + m) + \frac{(3-m)(2-m)^2}
{(1-m)^2}b\left(\frac{T_2}{T_1}\right)^2 - 3 \beta_0^2\right],   
\end{equation}
\[
32 \pi (p - \xi \theta)  = e^{-A}\Big[a(1+m)\left(\frac{Z_1}{Z_2}\right)^2 + \frac{4b(2-m)}
{(1-m)} + \frac{b(2-m)(2m^2 -7m + 2)}{(1-m)^2}\times
\]
\begin{equation}
\label{eq26}
\left(\frac{T_2}{T_1}\right)^2 - 3 \beta_0^2\Big],   
\end{equation}
where \\
$
Z_1 = c_1 -  c_2 \tan(z\sqrt{a})\\
Z_2 = c_2 + c_1 \tan(z\sqrt{a})\\
T_1 = \bar{c}_2 + \bar{c}_1 \tan(t\sqrt{b})\\
T_2 = \bar{c}_1 + \bar{c}_2 \tan(t\sqrt{b})\\
$
Here $\xi$, in general, is a function of time. The expression for kinematical parameter 
expansion $\theta$ is given by
\begin{equation}
\label{eq27}  
\theta = \frac{e^{-A/2}}{(m - 1)}\left(\frac{T_{3}}{T_{1}}\right),
\end{equation}
where
$
T_{3} = \bar{c}_1 - \bar{c}_2 \tan(t\sqrt{b}).
$
Thus, given $\xi(t)$ we can solve Eq. (\ref{eq26}). In most of the investigations 
involving bulk viscosity is assumed to be a simple power function of the energy 
density\cite{ref49}$-$ \cite{ref52}
\begin{equation}
\label{eq28} 
\xi(t) = \xi_{0} \rho^{n},
\end{equation}
where $\xi_{0}$ and $n$ are constants. For small density, $n$ may even be equal to unity 
as used in Murphy's work for simplicity\cite{ref24}. If $n = 1$, Eq. (\ref{eq28}) may 
correspond to a radiative fluid\cite{ref53}. Near the big bang, $0 \leq n \leq \frac{1}{2}$ 
is a more appropriate assumption\cite{ref54} to obtain realistic models.  

For simplicity and realistic models of physical importance, we consider the following two 
cases $(n = 0, 1)$:

\subsubsection {Model I: ~ ~ solution for  $\xi = \xi_{0}$}
When $n = 0$, Eq. (\ref{eq28}) reduces to $\xi = \xi_{0}$ = constant. Hence in this case 
Eq. (\ref{eq26}), with the use of (\ref{eq27}), leads to
\[
32 \pi p = \frac{32\pi\xi_{0}e^{-A/2}}{(m - 1)}\left(\frac{T_{3}}{T_{1}}\right) + e^{-A}
\Big[a(1 + m)\left(\frac{Z_{1}}{Z_{2}}\right)^{2} + \frac{4b(2 - m)}{(1 - m)} 
\]
\begin{equation}
\label{eq29} 
+ \frac{b(2 - m)
(2m^{2} - 7m + 2)}{(1 - m)^{2}}\left(\frac{T_{2}}{T_{1}}\right)^{2} - 3\beta_{0}\Big].
\end{equation}

\subsubsection {Model II: ~ ~ solution for $\xi = \xi_{0} \rho$}
When $n = 1$, Eq. (\ref{eq28}) reduces to $\xi = \xi_{0} \rho $ and hence Eq. (\ref{eq26}),
with the use of (\ref{eq27}), leads to
\[
32 \pi p = e^{-A}\Biggl[a(1 + m)(1 + T_{4}) + 4aT_{4} + \frac{4b(2 - m)}{(1 - m)} + 
\frac{b(2 - m)}{(1 - m)^{2}}\left(\frac{T_{2}}{T_{1}}\right)^{2}\times
\]
\begin{equation}
\label{eq30} 
\left\{(2 - m)(3 - m)T_{4} + 2m^{2} - 7m +2\right\} - 3(T_{4} + 1){\beta_{0}}^{2}\Biggr],
\end{equation}
where
$T_{4} = \frac{32\pi\xi_{0}e^{-a/2}}{(m - 1)}\left(\frac{T_{3}}{T_{1}}\right)$.

From the above results in both models it is evident that at any instant the domain wall 
density $\rho$ and pressure $p$ in the perpendicular direction decreases on both 
sides of the wall away from the symmetry plane and both vanish as 
$z \longrightarrow \pm \infty $. The space times in both cases are reflection symmetry 
with respect to the wall. All these properties are very much expected for a domain wall.
It can be also seen that the viscosity, as well as the displacement field $\beta$ exhibit
essential influence on the character of the solutions.  

\subsection{Case II: Time varying displacement field ($\beta = \beta_{0} t^{\alpha}$)} 
Using the aforesaid power law relation between time coordinate and displacement field,
Eq. (\ref{eq19}) may be written as
\begin{equation}
\label{eq31}
\ddot{w} - \Big[\frac{3(1-m)\beta_{0}^{2}}{4(2 - m)^2}t^{2\alpha} - \frac{k (1 - m)}
{(2 - m)^{2}}\Big]w = 0,
\end{equation}
where
\begin{equation}
\label{eq32}
w = e^{-\frac{(1-m)}{(2-m)}} B_2.
\end{equation}
Now, it is difficult to find a general solution of Eq. (\ref{eq31}) and hence
we consider a particular case of physical interest. It is believed that $\beta^2$ 
appears to play the role of a variable cosmological term $\Lambda(t)$ in Einstein's equation.  
Considering $\alpha = - 1$, $\beta = \frac{\beta_0}{t}$, Eq. (\ref{eq31}) reduces to
\begin{equation}
\label{eq33}
t^2 \ddot{w} + \Big[\frac{k(1-m)}{(2-m)^2}t^2 - \frac{3}{4}\frac{(1-m)}{(2-m)^2} 
\beta_0^2\Big] w = 0.
\end{equation}
Eq. (\ref{eq33}) yields the general solution
\begin{equation}
\label{eq34}
w t^{r +1} = (t^{3} D)^{r}\Big[\frac{c_1 e^{ht} + c_2 e^{-ht}}{t^{2r-1}}\Big],
\end{equation}
where \\
$ D \equiv \frac{d}{dt}, \\
r = \frac{1}{2}[ \{1 + \frac{3(1-m)}{(2-m)^2} \beta_0^2\}^{\frac{1}{2}} - 1], \\
h^2 = \frac{k(1-m)}{(2-m)^2}.$

For $r = 1$, $\beta_0^2 = \frac{8(2-m)^2}{3(1-m)}$, Eq. (\ref{eq34}) suggests
\begin{equation}
\label{eq35}
w = \left(h -\frac{1}{t}\right)c_3 e^{ht} - \left(h + \frac{1}{t}\right) c_{4} e^{-ht},
\end{equation}
where $c_3$ and $c_4$ are integrating constants.

Hence the metric coefficients have the explicit forms when $a > 0$ as
\begin{equation}
\label{eq36}
e^{A} = \left[c_1 \sin(z\sqrt{a}) + c_2 \cos(z\sqrt{a})\right]^{m}\times \left[\left(h - \frac{1}
{t}\right) c_3 e^{ht} - \left(h + \frac{1}{t}\right) c_4 e^{-ht}\right]^{(m - 1)},  
\end{equation}
\begin{equation}
\label{eq37}
e^B = \left[c_1 \sin(z\sqrt{a}) + c_2 \cos(z\sqrt{a})\right]\times \left[\left(h - \frac{1}{t}\right)
c_3 e^{ht} - \left(h + \frac{1}{t}\right)c_4 e^{-ht}\right]^{-\frac{(2 - m)}{(1 - m)}}.
\end{equation}

With the help of Eqs. (\ref{eq36}) and (\ref{eq37}), the energy density and 
pressure can be obtained from Eqs. (\ref{eq6}) and (\ref{eq7})
\begin{equation}
\label{eq38}
32 \pi \rho = e^{-A}\left[4a + a(1 + m)\left(\frac{Z_1}{Z_2}\right)^2 + \frac{(3-m)(2-m)^2}
{(1-m)^2}\left(\frac{c_3 h^2 t}{T_6} - \frac{1}{t}\right)^2 - \frac{3 \beta_0^2}{t^2}\right],
\end{equation}
\[
32\pi(p - \xi \theta) = e^{-A}\Biggl[a(1+m)\left(\frac{Z_1}{Z_2}\right)^2 - \frac{(1 + 2m)
(2 - m)^2}{(1 - m)^2}\left(\frac{c_3 h^2 t}{T_6} - \frac{1}{t}\right)^2  
\]
\begin{equation}
\label{eq39}
- \frac{4(2 - m)}{(1 - m)}\left\{\frac{1}{t^2} - \frac{4 c_4 h^3 t e^{-2ht}}{T_6} 
+ h^2 \left(\frac {T_5}{T_6}\right)^{2}\right\} - \frac{3 \beta_{0}^{2}}{t^2}\Biggr],
\end{equation}
where \\
$ 
T_5 = c_3 + c_4 (1 + 2 h t) e^{-2ht}, \\
T_6 = c_3 (ht - 1) - c_4 (1 + ht) e^{-2ht}. \\
$ 

The expression for kinematical parameter expansion $\theta$ is given by
\begin{equation}
\label{eq40}  
\theta = \frac{(hT_{7} + T_{8})(m^{2} - 4m + 5)}{2(m - 1)}e^{-A/2},
\end{equation}
where \\
$
T_{7} = (ht - 1)c_{3} + (ht + 1)c_{4}e^{-2ht}, \\
T_{8} = c_{3} + c_{4}e^{-2ht}. \\
$
In this case we again consider the following two cases ($n = 0, 1$):

\subsubsection {Model I: ~ ~ solution for  $\xi = \xi_{0}$}
When $n = 0$, Eq. (\ref{eq28}) reduces to $\xi = \xi_{0} = constant$. Hence in this case 
Eq. (\ref{eq39}), with the use of (\ref{eq40}), leads to
\[
32 \pi p = \frac{32\pi\xi_{0}(hT_{7} + T_{8})(m^{2} - 4m + 5)}{2(m - 1)T_{6}}e^{-A/2} +
e^{-A}\Biggl[a(1 + m)\left(\frac{Z_{1}}{Z_{2}}\right)^{2} - 
\]
\[
\frac{(1 + 2m)(2 - m)^{2}}{(1- m)^{2}}
\left(\frac{c_{3}h^{2}t}{T_{6}} - \frac{1}{t}\right)^{2}
- \frac{4(2 - m)}{(1 - m)}\times
\]
\begin{equation}
\label{eq41} 
\left\{\frac{1}{t^{2}} - \frac{4c_{4}h^{3}te^{-2ht}}{T_{6}} + 
h^{2}\left(\frac{T_{5}}{T_{6}}\right)^{2}\right\} - \frac{3\beta_{0}^{2}}{t^{2}}\Biggr]
\end{equation}

\subsubsection {Model II: ~ ~ solution for  $\xi = \xi_{0} \rho$}
When $n = 1$, Eq. (\ref{eq28}) reduces to $\xi = \xi_{0} \rho$. Hence in this case 
Eq. (\ref{eq39}), with the use of (\ref{eq40}), leads to
\[
32 \pi p = e^{-A}\left[4a + a(1 + m)\left(\frac{Z_{1}}{Z_{2}}\right)^{2} + \frac{(3 - m)
(2 - m)^{2}}{(1- m)^{2}}\left(\frac{c_{3}h^{2}t}{T_{6}} - \frac{1}{t}\right)^{2} - 
\frac{3\beta_{0}^{2}}{t^{2}}\right]T_{9} 
\]
\[
+ e^{-A}\Biggl[a(1 + m)\left(\frac{Z_{1}}{Z_{2}}\right)^{2} - 
\frac{(1 + 2m)(2 - m)^{2}}{(1- m)^{2}}
\left(\frac{c_{3}h^{2}t}{T_{6}} - \frac{1}{t}\right)^{2}
- \frac{4(2 - m)}{(1 - m)}\times
\]
\begin{equation}
\label{eq41} 
\left\{\frac{1}{t^{2}} - \frac{4c_{4}h^{3}te^{-2ht}}{T_{6}} + 
h^{2}\left(\frac{T_{5}}{T_{6}}\right)^{2}\right\} - \frac{3\beta_{0}^{2}}{t^{2}}\Biggr],
\end{equation}
where \\ 
$
T_{9} = \frac{16\pi\xi_{0}(hT_{7} + T_{8})(m^{2} - 4m + 5)}{(m - 1)T_{6}}e^{-A/2}. \\
$
From the above results in both cases it is evident that at any instant the domain wall 
density $\rho$ and pressure $p$ in the perpendicular direction decreases on both 
sides of the wall away from the symmetry plane and both vanish as 
$z \longrightarrow \pm \infty $. The space times in both cases are reflection symmetry 
with respect to the wall. All these properties are very much expected for a domain wall.
It can be also seen that the viscosity, as well as the displacement field $\beta$ exhibit
essential influence on the character of the solutions.  

\section {Study of geodesics}
The trajectory of the test particle $x^{i}\{t(\lambda), x(\lambda), y(\lambda), z(\lambda)\}$
in the gravitational field of domain wall can be determined by integrating the geodesic 
equations
\begin{equation}
\label{eq43}
\frac{d^2 x^{\mu}}{d \lambda^2} + \Gamma^{\mu}_{\alpha \beta} \frac{dx^{\alpha}}
{d\lambda}\frac{dx^{\beta}}{d\lambda} = 0,
\end{equation}
\noindent
for the metric (\ref{eq6}). It has been already mentioned in \cite{ref37}, the acceleration 
of the test particle in the direction perpendicular to the domain wall ( i.e. in 
the z-direction) may be expressed as
\begin{equation}
\label{eq44}
\ddot{z} = \frac{e^{B-A}}{2} \frac{\partial B}{\partial z} (\dot{x}^2 + \dot{y}^2)
- \frac{1}{2}\frac{\partial A}{\partial z}( \dot{t}^2 + \dot{z}^2) - 
\frac{\partial A}{\partial z} \dot{t} \dot{z}.
\end{equation}
\noindent
By simple but lengthy calculation one can get expression for acceleration
which may be positive, negative (or zero) depending on suitable choice of the constants.
This implies that the gravitational field of domain wall may be repulsive or attractive
in nature (or no gravitational effect). 
\section {Conclusions}
The present study deals with plane symmetric domain wall within the framework of 
Lyra geometry in presence of bulk viscous fluid. The essential difference between 
the cosmological theories based on Lyra geometry and Riemannian geometry lies in 
the fact that the constant vector displacement field $\beta$ arises naturally from 
the concept of gauge in Lyra geometry whereas the cosmological constant $\Lambda$ 
was introduced in {\it ad hoc} fashion in the usual treatment. Currently the study of 
domain walls and cosmological constant have gained renewed interest due to their 
application in structure formation in the universe. Recently Rahaman {\it et al.}\cite{ref37} 
have presented a cosmological model for domain wall in Lyra geometry under a specific 
condition by taking displacement fields $\beta$ as constant. The cosmological models 
based on varying displacement vector field $\beta$ have widely been considered in the 
literature in different contexts {\cite{ref32}$-$\cite{ref36}}. Motivated by these studies, 
it is worthwhile to consider domain walls with a time varying $\beta$ in Lyra geometry. 
In this paper both cases viz., constant and time varying displacement field $\beta$, are 
discussed in the context of domain walls with the framework of Lyra geometry. 

The study on domain walls in this paper successfully describes the various features of
the universe. A network of domain walls would accelerate the expansion of the universe, 
but it would also exert a repulsive force expected to help the formation of large-scale 
structures. An interesting result that emerged in this work is that the pressure 
perpendicular to the wall is non-zero. 

The effect of bulk viscosity is to produce a change in perfect fluid and 
hence exhibit essential influence on the character of the solution. We observe here 
that Murphy's conclusion\cite{ref24} about the absence of a big bang type singularity 
in the infinite past in models with bulk viscous fluid, in general, is not true. The 
results obtained in\cite{ref20} also show that, it is, in general, not valid, since 
for some cases big bang singularity occurs in finite past.   
\section*{Acknowledgements}
\noindent
The authors (A. Pradhan and S. Otarod) would like to thank the Inter-University
Centre for Astronomy and Astrophysics, Pune, India for providing facility where part 
of this work was carried out. S. Otarod also thanks the Yasouj University for providing 
leave during this visit. The authors are grateful to the referee for his comments and 
suggestions to bring the paper in the present form.
\newline
\newline

\end{document}